\documentclass[12pt]{revtex4}
\usepackage{amssymb,epsf}
\begin{document}
\title{Soliton Decay in Coupled System of Scalar Fields}
\author{N. Riazi}\email{riazi@physics.susc.ac.ir}
\author{A. Azizi}\email{azizi@physics.susc.ac.ir}
\author{S.M. Zebarjad}\email{zebarjad@physics.susc.ac.ir}
\address{Physics Department and Biruni  Observatory,
         Shiraz University, Shiraz 71454, Iran}
\address{Institute for Studies in Theoretical Physics and
               Mathematics (IPM), Tehran 19395, Iran.}
\begin{abstract}
A system of coupled scalar fields is introduced which possesses a
spectrum of massive single-soliton solutions. Some of these
solutions are unstable and decay into lower mass stable solitons.
Some properties of the solutions are obtained using general
principles including conservations of energy and topological
charges. Rest energies are calculated via a variational scheme,
and the dynamics of the coupled fields are obtained by solving the
field equations numerically.
\end{abstract}
\maketitle
\section{Introduction}
Relativistic solitons, including those of the conventional
sine-Gordon equation, exhibit remarkable similarities with
classical particles. They exert short range forces on each other
and make collisions, without losing their identities \cite{a1}.
They are localized objects and do not disperse while propagating
in the medium. Because of their wave nature, they do tunnel a
barrier in certain cases, although this tunnelling is different
from the well-known quantum version \cite{a2}. Topological
solitons are stable, due to  the boundary conditions at spatial
infinity. Their existence, therefore, is essentially dependent on
the presence of degenerate vacua \cite{a33}.
\par
Topology provides an elegant way of classifying solitons in
various sectors according to the mappings between the degenerate
vacua of the field and the points at spatial infinity. For the
sine-Gordon system in 1+1 dimensions, these mapping are between
$\phi= 2n\pi$, $n\in \mathbb{Z}$ and $x=\pm \infty$, which
correspond to kinks and anti-kinks of the SG system. More
complicated mappings occur in solitons in higher dimensions. For
example, in cosmic strings, the vacuum is $S^1$ and topological
sectors corresponds to distinct mappings between this $S^1$ and a
large circle around the string. This leads to the homotopy group
$\pi^1(S^1)=\mathbb{Z}$.
\par
Coupled systems of scalar fields have been investigated by many
authors \cite{bas,riz2,a4}. Bazeia et al. \cite{bas} considered a
system of two coupled real scalar fields with a particular
self-interaction potential such that the static solutions are
derivable from first order coupled differential equations. Riazi
et al. \cite{riz2} employed the same method to investigate the
stability of the single soliton solutions of a particular system
of this type. Inspired by the well-known properties of sine-Gordon
equation, we introduce a coupled system of two real scalar fields
which shows a considerably reacher structure and dynamics. The
present system  is not of the form investigated in \cite{bas} and
\cite{riz2} and static solutions are not derivable from first
order differential equations.
\par
In our proposed system, a spectrum of solitons with different rest
energies exist which are stable, unstable or meta-stable,
depending on their energies and boundary conditions. Some of the
more massive solitons decay spontaneously into stable ones which
subsequently leave the interaction area. Note that the term
``soliton'' is used throughout  this paper for localized
solutions. The problem of integrability of the model is not
addressed here. Such non-dispersive solutions are called ``lumps''
by Coleman \cite{a5} to avoid confusion with true solitons of
integrable models. However, it has now become popular to use the
term soliton in its general sense.
\par
Pogosian \cite{po1} investigated kink solutions in bi-dimensional
$SU(N)\times Z_2$ models. He found that heavier  kinks tend to
break up into lighter ones. Comparing our results with those
reported by Pogosian, it is interesting to note that fairly
similar phenomena are observed in quite different systems. In an
earlier paper, Pogosian and Vachaspati \cite{po2} reported
$(N+1)/2$ distinct classes of kink solutions in an $SU(N)\times
Z_2$ field theory.
\par
The organization of the paper is as follows: In section
\ref{sec2}, we introduce the lagrangian density, dynamical
equations, and conserved currents of the proposed model. In
section \ref{sec3}, some exact solutions, together with the
corresponding charges and energies are derived. The necessary
nomenclature, and   general behavior of the solutions due to the
boundary conditions  are also  introduced in this section.
Numerical solutions corresponding to different boundary conditions
are presented, and properties of these solutions like their
charges, masses and their stability status are addressed in this
section. In order to investigate the stability of the numerical
solutions, their evolution are worked out numerically. The
dynamical evolution of the unstable solutions are investigated
further in section \ref{sec4}. Our final conclusion and a summary
of the results is given in the last section.
\section{ Dynamical Equations and Conserved Currents \label{sec2}}
Our choice of the Lagrangian density reads
\begin{equation} \label{lag}
{\cal L}= \frac{1}{2}\partial^\mu \phi_1 \partial_\mu \phi_1 -
  \alpha_1({\phi_2}^2 + \epsilon_1)(1-cos\phi_1)+
  \frac{1}{2}\partial^\mu \phi_2 \partial_\mu \phi_2 -
  \alpha_2({\phi_1}^2 + \epsilon_2)(1-cos\phi_2)
\end{equation}
in which $\alpha_{1,2}$ and  $\epsilon_{1,2}$ are positive
constants, and $\phi_{1,2}$ are two real scalar fields. The
background space-time is assumed to have the metric $g_{\mu
\nu}=\mathrm{diag}(1,-1)$ in 1+1 dimensions and $c=1$ has been
used throughout this paper. Recall that the sine-Gordon Lagrangian
density
\begin{equation}
{\cal L}= \frac{1}{2}\partial^\mu \phi \partial_\mu \phi
-\alpha(1-\cos\phi)
\end{equation}
leads to single soliton solutions
\begin{equation}
\phi= 2n\pi  \pm 4\tan^{-1} e^{\sqrt{\alpha}x}
\end{equation}
all having rest energies $8\sqrt{\alpha}$. Our proposed Lagrangian
density (\ref{lag}) comes from the idea of  mixing two scalar
fields in such a way that the discrete vacua at $ \phi_1=2 n\pi$,
$\phi_2=2m\pi$ ($m,n\in \mathbb{Z}$) survive, while the rest
energies of solitons for each field (say $\phi_1$) is affected by
the value of the other ($\phi_2$) and vice versa. It will be seen
later, that this idea leads to the appearance of non-degenerate
solitons with distinct topological charges.
\par
The corresponding equations of motion are easily obtained by
applying the variational principle $\delta \big(\int {\cal L}
d^2x\big)=0$ to the lagrangian density (\ref{lag}):
\begin{equation}  \label{phi1}
  \square \phi_1 =\alpha_1
  ({\phi_2}^2+\epsilon_1)\sin\phi_1+2\alpha_2\phi_1(1-\cos\phi_2),
\end{equation}
and
\begin{equation} \label{phi2}
  \square \phi_2 =\alpha_2
  ({\phi_1}^2+\epsilon_2)\sin\phi_2+2\alpha_1\phi_2(1-\cos\phi_1).
\end{equation}
Since the lagrangian density (\ref{lag}) is Lorentz invariant, the
corresponding  energy-momentum tensor \cite{a3}
\begin{equation}
  T^{\mu \nu}=\partial^\mu \phi_1 \partial^\nu \phi_1 +
  \partial^\mu \phi_2 \partial^\nu \phi_2 -g^{\mu \nu} {\cal L},
  \label{tmunu}
\end{equation}
satisfies the conservation  law
\begin{equation}
  \partial_\mu T^{\mu \nu}=0.
\end{equation}
 The Hamiltonian density is obtained from Eq. (\ref{tmunu}) according to
\begin{equation}
  {\cal H}
  =T^0_0=\frac{1}{2}\dot{\phi_1}^2+\frac{1}{2}\dot{\phi_2}^2+
  \frac{1}{2}{{\phi'}_1}^2+\frac{1}{2}{{\phi'}_2}^2+V(\phi_1,\phi_2),
\end{equation}
where dot and prime denote time and space derivatives,
respectively. It is seen from (\ref{lag}) that the potential is
\begin{equation}
  V(\phi_1,\phi_2 )=\alpha_1({\phi_2}^2 +\epsilon_1)(1 - \cos\phi_1)+
  \alpha_2({\phi_1}^2 +\epsilon_2)(1 - \cos\phi_2).
\end{equation}
\begin{figure}[t]
  \epsfxsize=10cm
  \centerline{\epsffile{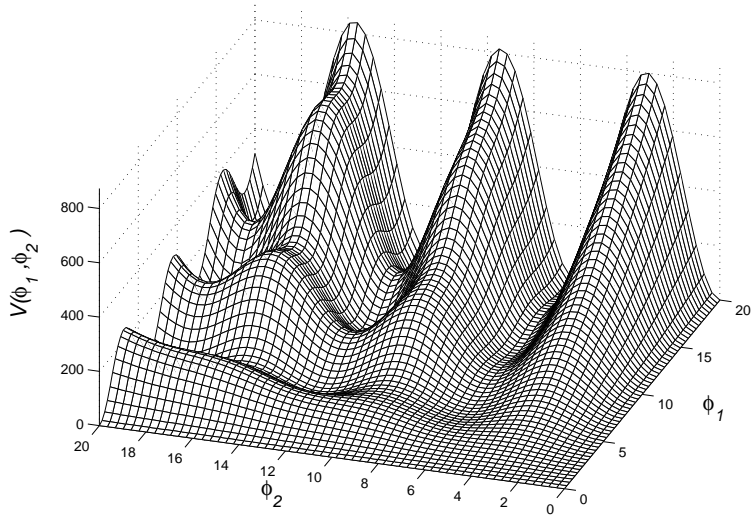}}
  \caption{ The self-interaction potential is shown as a height diagram
      over the $(\phi_1 , \phi_2 )$ plane.}
  \label{potential}
\end{figure}
This potential is shown in Fig.~(\ref{potential}), as a height
diagram over the $(\phi_1 ,\phi_2)$ plane. In this figure and any
numerical calculation throughout this paper, we fix the parameters
as $\alpha_1=0.3$, $\alpha_2=1$, $\epsilon_1=0.5$, and
$\epsilon_2=0.07$ for the sake of being definite. The classical
vacua consist of points in the $(\phi_1 ,\phi_2)$ plane at which
the condition $V(\phi_1, \phi_2)=0$ holds. Since the  potential
consists of two non-negative terms, it vanishes if and only if the
two terms vanish simultaneously. This leads to the vacuum set of
points in the $(\phi_1 ,\phi_2)$ space:
\begin{eqnarray} \label{calv}
  {\cal V}=\{(\phi_1,\phi_2)|\phi_1=2m\pi \ {\rm ~and~}  \
  \phi_2=2n\pi ;\ \ m,n\in \mathbb{Z}\}.
\end{eqnarray}
 It can be shown that the following
topological currents can be defined, which are conserved
independently, and lead to quantized charges:
\begin{eqnarray}
J^{\mu}_{mH} &=& \delta_{m p} \epsilon^{\mu \nu} \partial_\nu
                 \phi_1/2\pi \nonumber \\
J^{\mu}_{mV} &=& \delta_{m q} \epsilon^{\mu \nu}
  \partial_\nu\phi_2 / 2\pi. \label{cur}
\end{eqnarray}
In these equations, $m$ is an integer, and the integers $p$ and
$q$ are defined  according to
\begin{eqnarray}
  p &=& {\rm ~integer~part~of~} (\phi_{1}/2\pi) +1, \nonumber \\
  q &=& {\rm ~integer~part~of~} (\phi_{2}/2\pi) +1. \nonumber
\end{eqnarray}
The subscripts $V$ and $H$ denote ``Vertical'' and ``Horizontal''
which will be explained later. The currents $J^\mu_{mH,V}$ are
 conserved, independent of each other:
\begin{eqnarray}
  \partial_\mu J^\mu_{mH} &=& 0, \nonumber \\
  \partial_\mu J^\mu_{mV} &=& 0.
\end{eqnarray}
The corresponding topological charges are given by
\begin{eqnarray}
  Q_{mH} &=& \int_{-\infty}^\infty J^0_{mH} dx\; =\;
  \delta_{mp}\left[\phi_1(+\infty) - \phi_1(-\infty)\right] /
  2\pi,
  \nonumber \\
  Q_{mV} &=& \int_{-\infty}^\infty J^0_{mV} dx\; =\;
  \delta_{mq}\left[\phi_2(+\infty) - \phi_2(-\infty)\right] /
  2\pi.
\end{eqnarray}
These charges quantify  the mappings between the vacua
$\phi_{1,2}(\pm\infty)\in \cal{V}$ and the points at spatial
infinity.
\section{Single Soliton Solutions \label{sec3}}
\begin{figure}[t]
\epsfxsize=5.5cm
  \centerline{\epsffile{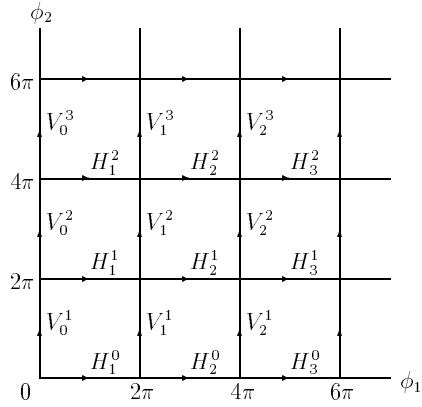}}
  \caption{Nomenclature of horizontal ($H$) and
    vertical ($V$) solutions according to boundary conditions. The
three lightest solutions: $H_1^0$, $V_0^1$, and $H_1^1$ are
stable, while $V_1^1$ and $V_2^1$ are unstable. The corresponding
rest energies and charges are shown in Table \ref{tab1} for our
choice of parameters.}
  \label{phi1phi2}
\end{figure}

\begin{figure}[t]
\epsfxsize=10cm
  \centerline{\epsffile{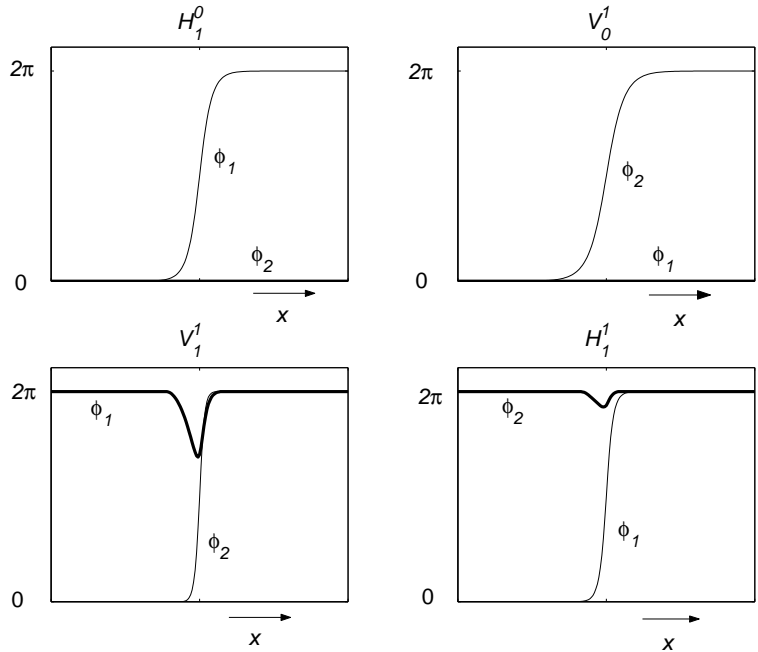}}
  \caption{Examples of static solutions.}
  \label{statsol}
\end{figure}

\begin{figure}
\epsfxsize=7cm
  \centerline{\epsffile{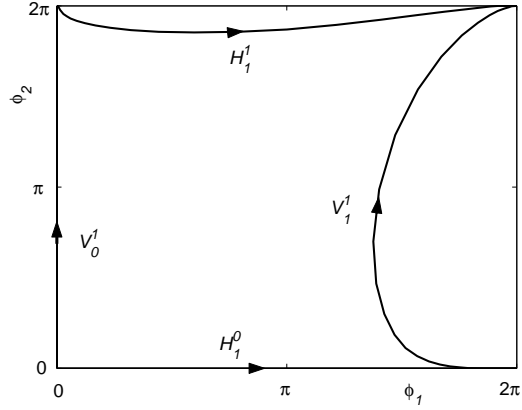}}
  \caption{The static solutions $V^1_0$, $H^0_1$, $H^1_1$, and $V^1_1$
  as projected on the ($\phi_1 ,\phi_2$) plane.}
  \label{solmap}
\end{figure}

Static solutions which correspond to transitions between adjacent
vacua are symbolically shown in Fig. (\ref{phi1phi2}).
Accordingly, we call the static solutions $H$ (horizontal) and $V$
(vertical) types. We have called them `horizontal'
 and `vertical' simply because of their orientation in
($\phi_1, \phi_2$) plane. Initial guesses for these solutions are
obtained using the known properties of the conventional
sine-Gordon equation \cite{a4}. Note that the exact solutions are
not strictly horizontal or vertical lines in the ($\phi_1,
\phi_2$) plane. Rather, they are bent curves joining two
neighboring vacuum points due to the coupling between the two
scalar fields. Any finite energy solution should start and end at
one of the vacuum points belonging to $\cal{V}$ (Eq. \ref{calv}).
 Equations (\ref{phi1}) and (\ref{phi2}) possess the following
 exact single soliton solutions:
\begin{equation} \label{sphi1}
  H^0_{p+1}: \qquad
  \phi_1=4\tan^{-1}\exp\left[ \pm a (x-x_0) \right] +2p\pi
  \quad{\rm ~and~}\quad \phi_2=0,
\end{equation}
with
\[
  Q_{mH} = \delta_{mp}, \qquad Q_{mV} = 0,
\]
and
\begin{equation}\label{sphi2}
  V_0^{q+1}: \qquad\ \ \quad
  \phi_2=4\tan^{-1}\exp\left[ \pm b (x-x_0) \right] +2q\pi
  \quad{\rm ~and~}\quad \phi_1=0,
\end{equation}
with \[
  Q_{mH} = 0, \qquad Q_{mV} = \delta_{mq},
\]
where $p$ and $q$ are integers. Note that Eqs. (\ref{sphi1}) and
(\ref{sphi2}) satisfy Eqs. (\ref{phi1}) and (\ref{phi2})
 in the static ($\frac{\partial}{\partial t}=0$) case.
In these equations, $a=\sqrt{\alpha_1\epsilon_1}$,
$b=\sqrt{\alpha_2\epsilon_2}$, and $x_0$ is the kink position.
Apart from these exact solutions, we will introduce other static
solutions which are obtained numerically later. Despite the
similarity of the special solutions (\ref{sphi1}) and
(\ref{sphi2}) with those of sine-Gordon equation, there are
profound differences between the general static solutions of the
present system and SG, including non-degenerate soliton masses and
instability of some of the static solutions. The rest energies of
the $H$ and $V$-type solitons are obtained by integrating the
corresponding Hamiltonian densities (with
$\dot{\phi_1}=\dot{\phi_2}=0$) over the $x$-space. The rest
energies are approximately given by:
\begin{equation}\label{hmass}
H-{\rm type}: \ \ M_H\simeq 8\sqrt{\alpha_1(4\pi^2 n^2
+\epsilon_1)},
\end{equation}
and
\begin{equation}\label{vmass}
V-{\rm type}: \ \ M_V\simeq 8\sqrt{\alpha_2(4\pi^2
n^2+\epsilon_2)},
\end{equation}
where $n$ is an integer. We have written  programs in  the
`MATLAB' environment, which calculate static solutions and the
dynamical evolutions from  pre-specified initial conditions. The
static solutions are obtained by minimizing the energy functional
\begin{equation}
  E = \int {\cal H} dx
\end{equation}
where ${\cal H}=T_0^0$ is the energy (Hamiltonian) density, using
a variational procedure.   The program is asked to terminate as
soon as the energy converges up to a $10^{-6}$ accuracy. Inspired
by the exact solutions (\ref{sphi1}) and (\ref{sphi2}), the
initial guesses were taken to be of the form of the sine-Gordon
solitons. These initial guesses were deformed toward the lowest
energy solutions by the variational calculation. This deformation
is caused by the slope of the potential function as depicted in
Fig. (\ref{potential}). Sample solutions are shown in Fig.
(\ref{statsol}).
Static solutions $V^1_0$, $H^0_1$, $H^1_1$ and $V^1_1$  as
projected on the ($\phi_1$,$\phi_2$) plane are shown in Fig.
(\ref{solmap}).
A sample classification is shown in Table \ref{tab1} for a
particular choice of the parameters.
\begin{table}
\begin{center}
\begin{tabular}{ccccccccccccccl} \hline
symbol && mass && $Q_{1H}$ && $Q_{1V}$ && $Q_{2H}$ &&  $Q_{2V}$
       && $Q_{3H}$ &&stability \\
\hline
$V^1_0$ &&2.09 && 0  && +1 && 0  &&  0 && 0 &&stable \\
$V^2_0$ &&2.09 && 0  && 0  && 0  && +1 && 0 &&stable \\
$H^0_1$ &&3.09 && +1 && 0  && 0  &&  0 && 0 &&stable \\
$H^0_2$ &&3.09 && 0  && 0  && +1 &&  0 && 0 &&stable \\
$H^0_3$ &&3.09 && 0  && 0  &&  0 &&  0 && +1 &&stable \\
$H^1_1$ &&26.7 && +1 && 0  && 0  &&  0 && 0  &&stable \\
$H^1_2$ &&27.5 && 0  && 0  && +1 &&  0 && 0  &&stable \\
$H^1_3$ &&27.6 && 0  && 0  && 0  &&  0 && +1 &&stable \\
$V^1_1$ &&42.9 && 0  && +1 && 0  &&  0 && 0  &&
              $V^1_1\rightarrow \bar{H}^0_1V^1_0H^1_1$\\
$V^2_1$ &&48.4 && 0  && 0  && 0  && +1 && 0  &&stable \\
$H^2_1$ &&53.6 && +1 && 0  && 0  &&  0 && 0  &&stable \\
$H^2_2$ &&54.6 && 0  && 0  && +1 &&  0 && 0  &&stable \\
$V^1_2$ &&89.6 && 0  && +1 && 0  &&  0 && 0  &&
  $V^1_2\rightarrow \bar{H}^0_2\bar{H}^0_1V^1_0H^1_1H^1_2$\\
$V^2_2$ &&95.5 && 0  && 0  && 0  && +1 && 0  &&stable \\
$V^1_3$ &&129  && 0  && +1 && 0  &&  0 && 0  &&metastable \\
\hline
\end{tabular}
\caption{A sample classification of the lowest energy solitons for
         $\alpha_1=0.3$, $\alpha_2=1$, $\epsilon_1=0.5$,
         $\epsilon_2=0.07$.} \label{tab1}
\end{center}
\end{table}
Anti-solitons exist with the same masses (rest energies) as
solitons but with opposite charges. The anti-soliton solutions are
obtained by simply exchanging the boundary conditions at
$x=+\infty$ into those at $x=-\infty$. For example,
\begin{equation}
  H_1^0 : \left\{
  \begin{array}{ll}
    \phi_1(-\infty) = 0, & \phi_1(+\infty) = 2\pi \\
    \phi_2(-\infty) = 0, & \phi_2(+\infty) = 0,
  \end{array}
  \right.
\end{equation}
while
\begin{equation}
  \bar{H}_1^0 : \left\{
  \begin{array}{ll}
    \phi_1(-\infty) = 2\pi, & \phi_1(+\infty) = 0 \\
    \phi_2(-\infty) = 0, & \phi_2(+\infty) = 0.
  \end{array}
  \right.
\end{equation}
It is clear from the definition of topological charges (\ref{cur})
that solitons and anti-solitons have opposite charges. The
approximate formulae (\ref{hmass}) and (\ref{vmass}) may be
compared with the numerical results depicted in Table \ref{tab1}.
It can be seen that the masses of the $H$-type solitons are better
approximated by Eq. (\ref{hmass}) than those given by Eq.
(\ref{vmass}). This difference can be attributed to the fact that
the horizontal solitons deviate less from the sine-Gordon solitons
(see Fig.(\ref{solmap})).
\section{Soliton Decay \label{sec4}}
\begin{figure}
\epsfxsize=15cm
  \centerline{\epsffile{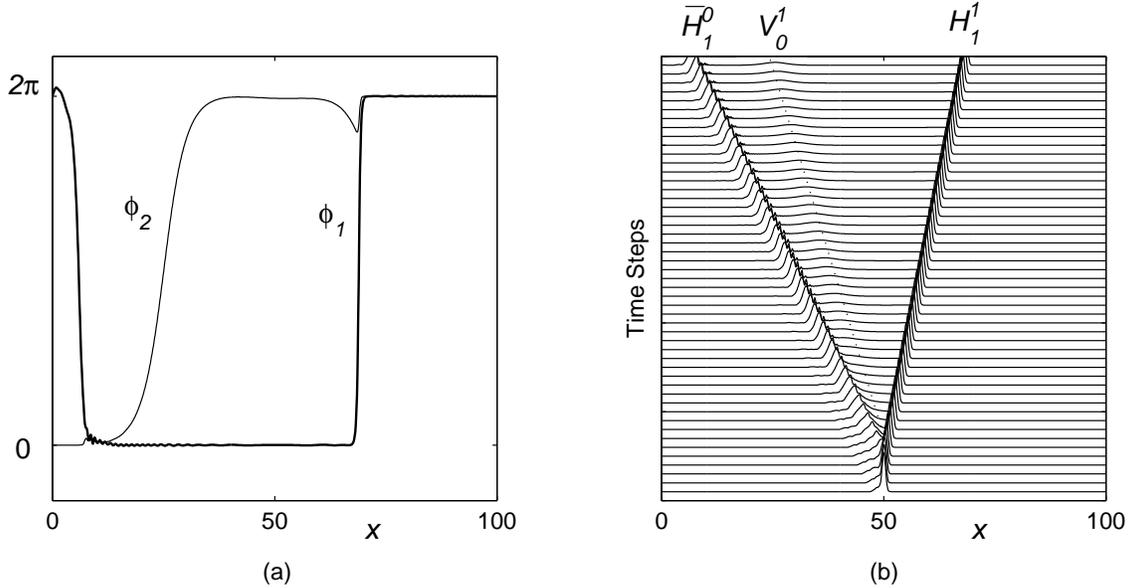}}
  \caption{The decay of $V^1_1$. (a) $\phi_1$ and $\phi_2$ as a
function of $x$ after the decay of $V^1_1$. (b) The energy density
as a function of $x$ for various time steps. The dotted line for
$V^1_0$ is added in order to clarify the trajectory of this decay
product.}
  \label{v11dyn}
\end{figure}

The time-dependent solutions of the dynamical equations
(\ref{phi1}) and (\ref{phi2}) were obtained numerically, by
transforming these PDEs  into finite difference equations, and
calculating the fields $\phi_1(x,t)$ and $\phi_2(x,t)$ in
successive time steps. Most of the static solutions obtained did
not undergo appreciable variations when inserted into our
dynamical program as initial conditions. This is a numerical
indication for the stability of the corresponding static
solutions. However, not all the static solutions obtained by the
variational method explained in the last section were found to be
stable. For the given choice of parameters, $V^1_1$, for example,
is unstable and decays spontaneously via (see Fig. \ref{v11dyn}(a)
and (b))
\begin{equation} \label{hvh}
  V^1_1 \longrightarrow \bar{H}^0_1 +V^1_0+ H^1_1.
\end{equation}
Here, $\bar{H}^0_1$ is the anti-soliton of $H^0_1$. Figure
(\ref{v11dyn}(a)) shows the $\phi_1$ and $\phi_2$ fields after
several time steps. The decay product solitons can be identified
at positions where the fields jump from one vacuum to an adjacent
vacuum. Figure (\ref{statsol}) helps recognizing these decay
products. Figure (\ref{solmap}) shows the stable solutions
$H^1_1$, $V^1_0$, and $H^0_1$ and the unstable solution $V_1^1$ on
the $(\phi_1,\phi_2)$ plane. Numerical calculations show that the
decay of $V^1_1$ starts with the trajectory shown for this
solution and evolves gradually to the $\bar{H^0_1}$ ($H^0_1$ in
the reverse direction), $V_0^1$ and $H^1_1$ trajectories in this
figure. Figure (\ref{v11dyn}(b)) shows the corresponding energy
density for various time steps. Individual solitons as decay
products are more apparent in this figure.
\par
The balance of $H$ and $V$ charges is of course respected in the
decay (\ref{hvh}), which can be easily demonstrated by computing
$Q_{mV}$ and $Q_{mH}$ for $V_1^1$, $\bar{H}^0_1$, $V_0^1$, and
$H_1^1$. From the conservation of energy point of view,
$M(V^1_1)=42.9$, while $M(\bar{H}^0_1)+M(V^1_0)+M(H^1_1)=31.88$
which shows that the decay is allowed, and the excess energy is
transferred to the kinetic energies  of the solitons. As Fig.
(\ref{v11dyn}(b)) shows, no appreciable energy is radiated away as
small amplitude wavelets.
\par
\begin{figure}
\epsfxsize=15cm
  \centerline{\epsffile{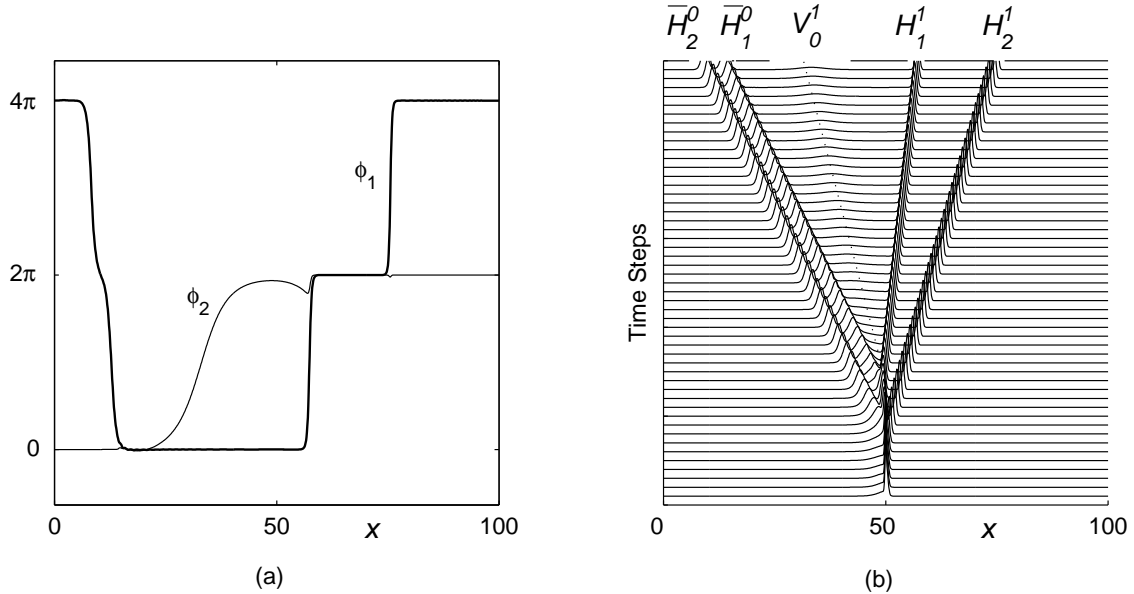}}
  \caption{The decay of $V^1_2$. (a) $\phi_1$ and $\phi_2$ as a
  function of $x$ after the decay of $V^1_2$. (b) The energy density
  as a function of $x$ for various time steps.}
  \label{v12dyn}
\end{figure}

 Numerical results show that $V^1_2 $ is also unstable and decays
 via
\begin{equation}\label{v12}
V^1_2 \longrightarrow \bar{H}^0_2 +\bar{H}^0_1+V^1_0+ H^1_1+H^1_2.
\end{equation}
Figure (\ref{v12dyn}) illustrates the decay  of $V^1_2$.
It can be seen from Fig. (\ref{v12dyn}(b)) that a
$\bar{H}^0_1V^1_0H^1_1$ compound (i.e. $V_1^1$) is produced first.
This compound subsequently decays into its components in a short
time. Note that $M(V^1_2)=89.6$ and the sum of the rest energies
of the decay products is $\sum_{\mathrm{product}} M=62.47$. The
difference goes to the kinetic energies of the decay products. The
total energy and all topological charges are conserved. In order
to demonstrate the conservation of topological charges, let us
write down the charges of individual solitons:
\begin{eqnarray}
V^1_2&:&~ Q_{1V}=+1 ~~~ \bar{H}^0_2:~ Q_{2H}=-1 ~~~  \bar{H}^0_1:~
    Q_{1H}=-1  \nonumber\\
V^1_0&:&~ Q_{1V}=+1 ~~~  H^1_1:~ Q_{1H}=+1~~~H^1_2:~Q_{2H}=+1\nonumber
\end{eqnarray}
all other charges being zero. We thus have $Q_{1V}$(LHS)=+1 and
$Q_{1V}$(RHS)=+1, and all other charges equal to zero for both
sides.
\par
 Surprisingly enough, although the decay of $V^1_3$ via
\begin{equation}\label{v13}
    V^1_3 \longrightarrow \bar{H}^0_3+\bar{H}^0_2 +
    \bar{H}^0_1+V^1_0+ H^1_1+H^1_2+H^1_3
\end{equation}
is allowed by conservation of energy and charge, such a decay is
not observed. We thus conclude that $V^1_3$ is a metastable
soliton, and its fission needs some external trigger.
\section{Conclusion \label{sec5|}}
Although the Sine-Gordon equation is an integrable system, even
minor modifications of the equation, usually exploit its
integrability. We described in this paper and elsewhere
\cite{a2,a6}, that interesting and rich behavior may result by
suitable modifications of the sine-Gordon equation irrespective of
being integrable or not. Removal of mass degeneracy, soliton
confinement, and soliton decay are among such properties.
\par
For the system introduced in this paper, we presented a class of
exact, single soliton solutions in the particular case where the
system reduced to the sine-Gordon equation ($\phi_1=0$ or
$\phi_2=0$). Other classes of static solutions were computed
numerically using a variational algorithm. These static solutions
were then fed into a numerical program which computed the
dynamical evolution of the solution. We found that most of the
static solutions were stable, while a few underwent decay into
lower energy solitons. Topological charges where conserved
throughout these dynamical processes, in addition to the
conservation of energy and linear momentum which results from the
invariance of Lagrangian under space and time translation. In the
particular decay (\ref{v12}), two interesting effects were
observed: First, the decay did not start promptly. Rather, $V^1_2$
decayed into $\bar{H}_2^0V_1^1H_2^1$ first, and $V_1^1$ decayed
via (\ref{hvh}) in a later stage. Second, the ($\bar{H}^0_1,
\bar{H}^0_2$) pair apparently remained as a bound system in the
decay of $V^1_2$. In order to check whether $\bar{H}^0_1$ and
$\bar{H}^0_2$ do form a bound system, we did numerical
calculations (both variational and dynamical) for this system. It
turned out that the system split into $\bar{H}^0_1$ and
$\bar{H}^0_2$ which moved away from each other. We conclude that
$\bar{H}^0_1\bar{H}^0_2$ do not form a bound system.
\par
Finally, let us discuss briefly the kinematics of solitons in the
decay process. Since our system is relativistic,  principles of
conservation of energy and momentum should be applicable in their
relativistic form. Ignoring the energy and momentum radiated away
in the form of small amplitude wavelets, we may write
\begin{equation}
\label{energyeq} Mc^2=\sum^n_{i=1}\gamma_im_ic^2,
\end{equation}
and
\begin{equation}
\label{momeq} \sum_{i=1}^n\gamma_im_iv_i=0,
\end{equation}
for the decay
\begin{equation}
S\rightarrow s_1+s_2+...+s_n.
\end{equation}
In equations (\ref{energyeq}) and (\ref{momeq}), $M$ is the mass
of the $S$ soliton, $m_i$ are the masses of $s_i$ solitons,  $v_i$
are the corresponding velocities, and $\gamma_i=(1-v_i^2)^{-1/2}$.
The system of equations (\ref{energyeq}) and (\ref{momeq}) have a
unique solution for $v_i$, if $n=2$.  For $n>2$, we have only two
equations for $n$ unknowns and the equations do not have a unique
solution. However, the decay pattern and the distribution of
velocities among the decay products seem to be predetermined in
our numerical results (compare the decay pattern of $V_1^1$ in
Figures (\ref{v11dyn}) and (\ref{v12dyn})).  If we note that the
decay to the daughter solitons does not happen in a single stage,
but rather it proceeds in successive stages, it becomes clear why
the distribution of velocities among the decay products follow a
unique pattern. In each stage, an unstable soliton decays into
{\bf two} decay products which is actually observed numerically.
The velocities of the two decay products is unique, according to
(\ref{energyeq}) and (\ref{momeq}). One of the decay products, in
turn, decays into two solitons in later stage, and so on.
\acknowledgements
 Support of Research Council of  Shiraz University (grant
 79-SC-1379-C129) and IPM is gratefully acknowledged.

%
\end{document}